\documentclass[%
superscriptaddress,
reprint,
showpacs,
 amsmath,amssymb,
 aps,
pra,
xcolor
]{revtex4-1}

\usepackage{xcolor}
\usepackage{graphicx}
\usepackage{dcolumn}
\usepackage{bm}
\usepackage{amssymb}

\begin{document}

\title{
Radiative properties of rubidium atoms trapped in solid neon and  parahydrogen}

\author{David M. Lancaster}
\affiliation{Department of Physics, University of Nevada, Reno NV 89557, USA}
\author{Ugne Dargyte}
\affiliation{Department of Physics, University of Nevada, Reno NV 89557, USA}
\author{Sunil Upadhyay}
\affiliation{Department of Physics, University of Nevada, Reno NV 89557, USA}
\author{Jonathan D. Weinstein}
\email{weinstein@physics.unr.edu}
\homepage{http://www.weinsteinlab.org}
\affiliation{Department of Physics, University of Nevada, Reno NV 89557, USA}


\begin{abstract}
It is known from ensemble measurements that rubidium atoms trapped in solid parahydrogen have favorable properties for quantum sensing of magnetic fields.
To use a single rubidium atom as a quantum sensor requires a technique capable of efficiently measuring the internal state of a single atom, such as laser-induced fluorescence.
In this work we search for laser-induced fluorescence from ensembles of rubidium atoms trapped in solid parahydrogen and, separately, in solid neon. In parahydrogen we find no evidence of fluorescence over the range explored, and place upper limits on the radiative branching ratio. In neon, we observe laser induced fluorescence, measure the spectrum of the emitted light, and measure the excited state lifetime in the matrix. Bleaching of atoms from the excitation light is also reported.
\end{abstract}

\maketitle

\section{Introduction}

{
Recent experimental work has demonstrated that the spin states of alkali atoms trapped in solid parahydrogen can be controlled and measured via optical techniques; the ensemble spin-state-dependent contrast is 0.1 \cite{upadhyay2016longitudinal, PhysRevA.100.063419}. Ensemble spin dephasing times are as long as T$_2^* \sim 60~\mu$s \cite{PhysRevA.100.063419, PhysRevB.100.024106}. With dynamical decoupling, spin coherence times T$_2$ as long as 0.1~s are observed \cite{PhysRevLett.125.043601}. Spin state control and readout, along with long coherence times, are the key properties for atomic magnetometers and quantum sensors \cite{budker2007optical, RevModPhys.89.035002}, and these numbers are competitive with the state-of-the-art for ensemble measurements of electron spins in systems capable of high spin densities \cite{barry2020sensitivity}. 
}

By using a single atom as a microscopic magnetic field sensor, it should be possible to perform NMR and MRI measurements of single molecules co-trapped in the parahydogen \cite{PhysRevLett.125.043601}, similar to what  has been previously proposed for NV centers in diamond \cite{taylor2008high, PhysRevX.5.011001, staudacher2013nuclear, mamin2013nanoscale, sushkov2014magnetic}.

{
Rubidium atoms trapped in solid parahydrogen exhibit large broadening  of their optical transitions, with linewidths $> 10^5$ larger than their natural linewidth \cite{PhysRevA.100.063419}. This significantly reduces both the cross-section for light scattering and the dispersive interaction of the atom with light.
This is not an impediment for making absorptive measurements of  ensembles of atoms trapped in parahydrogen, as the large number of atoms compensates for the weak interaction of individual atoms \cite{PhysRevA.100.063419}. However, it would be expected to make detection of a single atom by absorptive \cite{streed2012absorption} or dispersive means \cite{buckley2010spin} extremely challenging.
Laser-induced fluorescence is a more practical means of detecting the spin states of single atoms trapped in cryogenic matrices, \emph{if} the trapped atoms have favorable emission properties \cite{gruber1997scanning}.
}

We note that
laser-induced fluorescence is also a key technique for experiments seeking to use matrix-isolated atoms for nuclear physics and particle physics experiments \cite{PhysRevC.99.065805, chambers2019imaging}.

In this work, we report measurements of laser-induced fluorescence (LIF) from rubidium atoms trapped in solid parahydrogen and neon matrices.







\section{Prior work}

To our knowledge LIF measurements of rubidium atoms trapped in either neon or parahydrogen have not been previously reported in the literature. However, LIF has previously been observed for other combinations of alkali atoms and noble-gas matrices: helium \cite{Weis1995, moroshkin2006spectroscopy}, neon \cite{fajardo1998solid}, argon \cite{gerhardt2012excitation, doi:10.1063/1.444487}, krypton \cite{gerhardt2012excitation}, and xenon \cite{gerhardt2012excitation}.
Excited state lifetimes in the matrix have been measured to be comparable to their gas-phase values, indicating that radiative decay is the dominant decay path in the solid \cite{doi:10.1063/1.444487}. 
For alkali atoms the emitted light is typically spectrally broadened, on the order of tens of nanometers, and redshifted on the order of tens to hundreds of nanometers \cite{doi:10.1063/1.444487, gerhardt2012excitation}.
{
For other atomic species in noble gas matrices, significantly narrower emission and absorption lines have been observed in the near-infrared \cite{PhysRevA.99.022505}.

The majority of work to date has focused on ensemble measurements. However, the nEXO collabortion has successfully used LIF to image single barium atoms trapped in solid xenon \cite{chambers2019imaging}. While the $^1S_0$ ground state of barium is not particularly well-suited to quantum sensing of magnetic fields, we expect that any matrix-isolated atom which exhibits favorable LIF properties could similarly be imaged at the single-atom level.
}





\section{LIF measurements: neon}


The experimental apparatus is as described in Ref. \cite{PhysRevA.100.063419}. Samples are grown by co-deposition of rubidium and the matrix gas onto a cryogenic sapphire window in vacuum.

\begin{figure}[ht]
    \begin{center}
    \includegraphics[width=\linewidth]{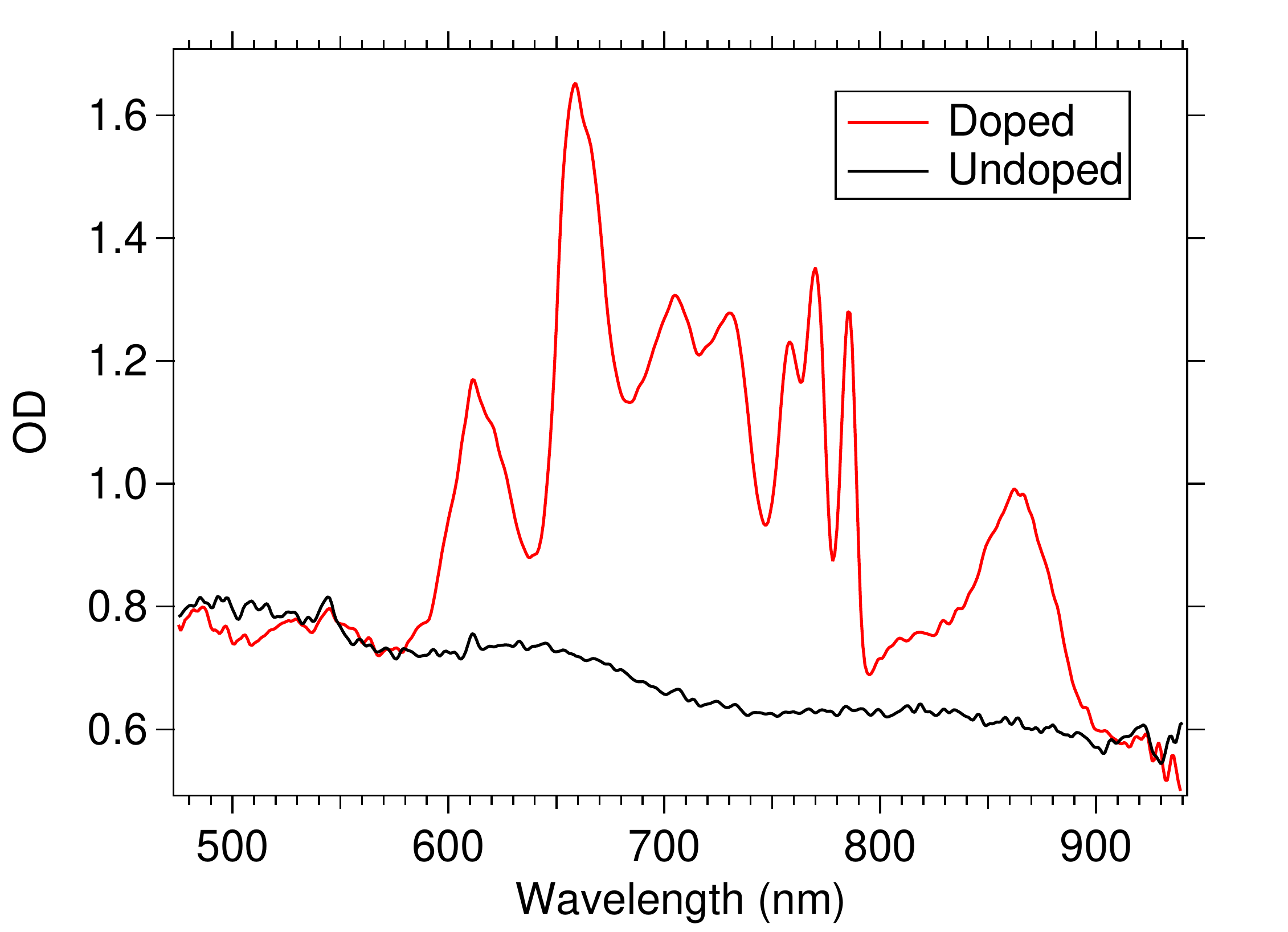}
    \caption{
Absorption spectra of rubidium-doped and undoped neon solids. The undoped spectrum has been scaled by a factor of the relative thicknesses of the doped and undoped samples (0.17 and 0.15~mm, respectively).
%
\label{fig:NeAbsSpectrum}
    }
    \end{center}
\end{figure}

White-light absorption spectra of both doped and undoped neon samples are shown in Fig. \ref{fig:NeAbsSpectrum}. 
We calculate the optical depth (OD) of the sample from the transmission $T$ with the definition $T \equiv e^{-OD}$. There is significant ``background'' scattering from the neon matrix itself, accompanied by multiple absorption lines from the implanted rubidium atoms. 
We attribute the different spectral peaks to a combination of different ``trapping sites'' in the matrix and the crystal field splitting in the excited state  \cite{PhysRevA.100.063419, PhysRev.166.207}.

The samples grown in Fig. \ref{fig:NeAbsSpectrum} were grown at a deposition rate of $\sim 0.6$~$\mu$m/minute, as measured by thin-film interferometry. The substrate temperature was $\sim 3.3$~K during sample growth, elevated from its base temperature of $2.9$~K by the heat load of sample growth. All subsequent measurements were performed at the base temperature. The central $\sim 0.1$~mm of the sample was doped with rubidium atoms. From the absorption spectrum, we calculate a Rb density of $\sim 3 \times 10^{17}$~cm$^{-3}$ in the doped region.



To search for laser-induced fluorescence, we excite our sample using linearly-polarized light from a narrow-linewidth tuneable Ti:sapphire laser. Typical excitation powers are $\sim 0.1$~mW, focused to a $\sim 0.3$~mm spot size on the sample. Emitted light from the sample is focused into a fiber-coupled grating spectrometer. 

\begin{figure}[ht]
    \begin{center}
    \includegraphics[width=\linewidth]{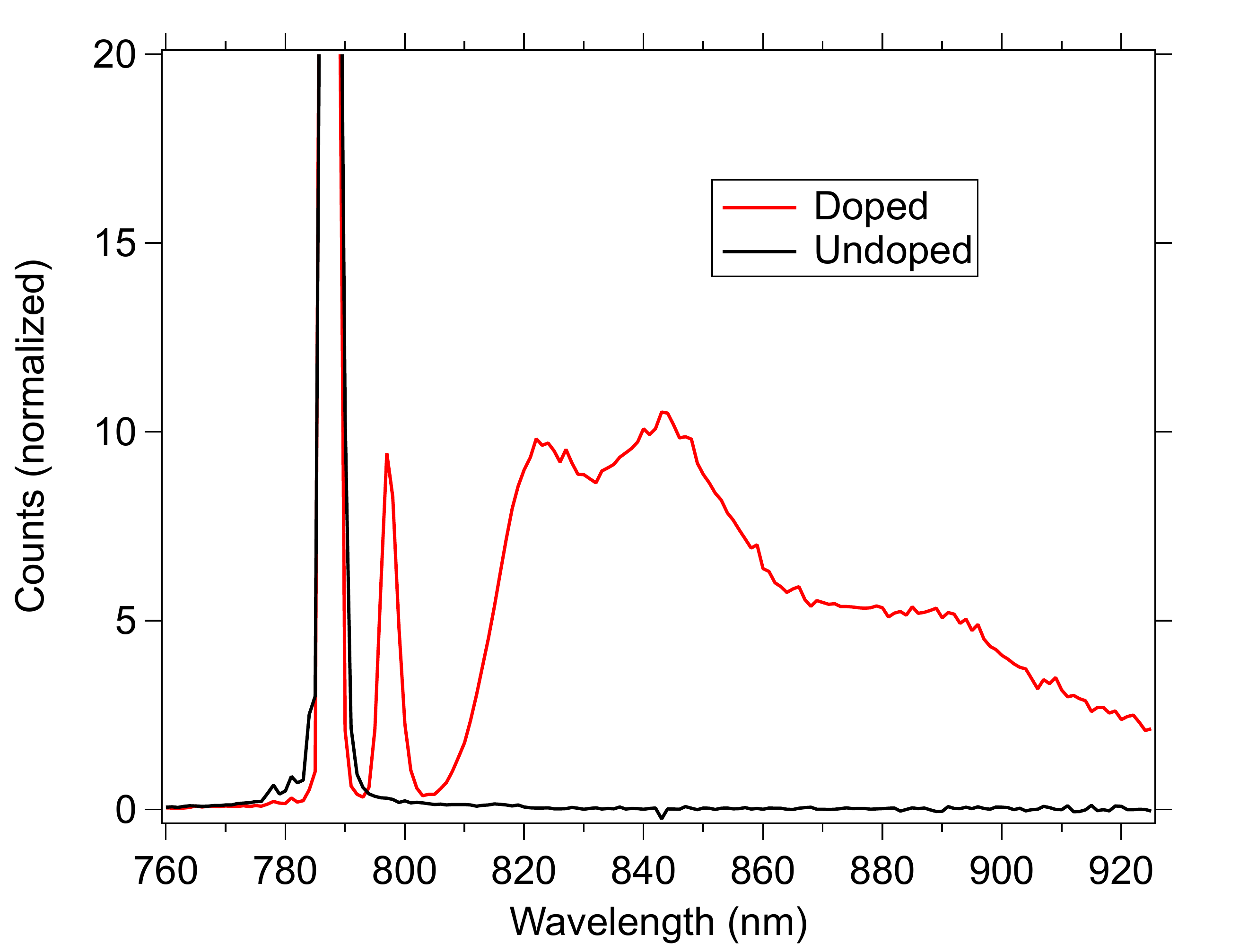}
    \caption{ 
LIF spectra of rubidium-doped and undoped neon solids for an excitation wavelength of 786~nm.  The samples are the same as in  Fig. \ref{fig:NeAbsSpectrum}. 
The spectrometer counts have been normalized by the excitation laser power and the spectral response of the detector; the resonant-scattering peak at the excitation wavelength is saturated. 
\label{fig:NeLIFspectrum}
    }
    \end{center}
\end{figure}

As seen in Fig. \ref{fig:NeLIFspectrum}, when exciting the narrow peak at 786~nm, we observe a large amount of elastically-scattered light, a narrow redshifted peak, and a broad peak at a larger redshift. By comparing LIF spectra of doped and undoped samples, we conclude the resonant light is predominantly from the neon while the redshifted light is predominantly rubidium LIF. To reduce the scatter from the neon we use a polarizer to block the resonant peak. We find this reduces the resonant scatter by over an order of magnitude, while reducing the redshifted LIF by only a factor of $\sim 2$, indicating the resonant scatter is nearly linearly polarized while the redshifted LIF is unpolarized.

\begin{figure}[ht]
    \begin{center}
    \includegraphics[width=\linewidth]{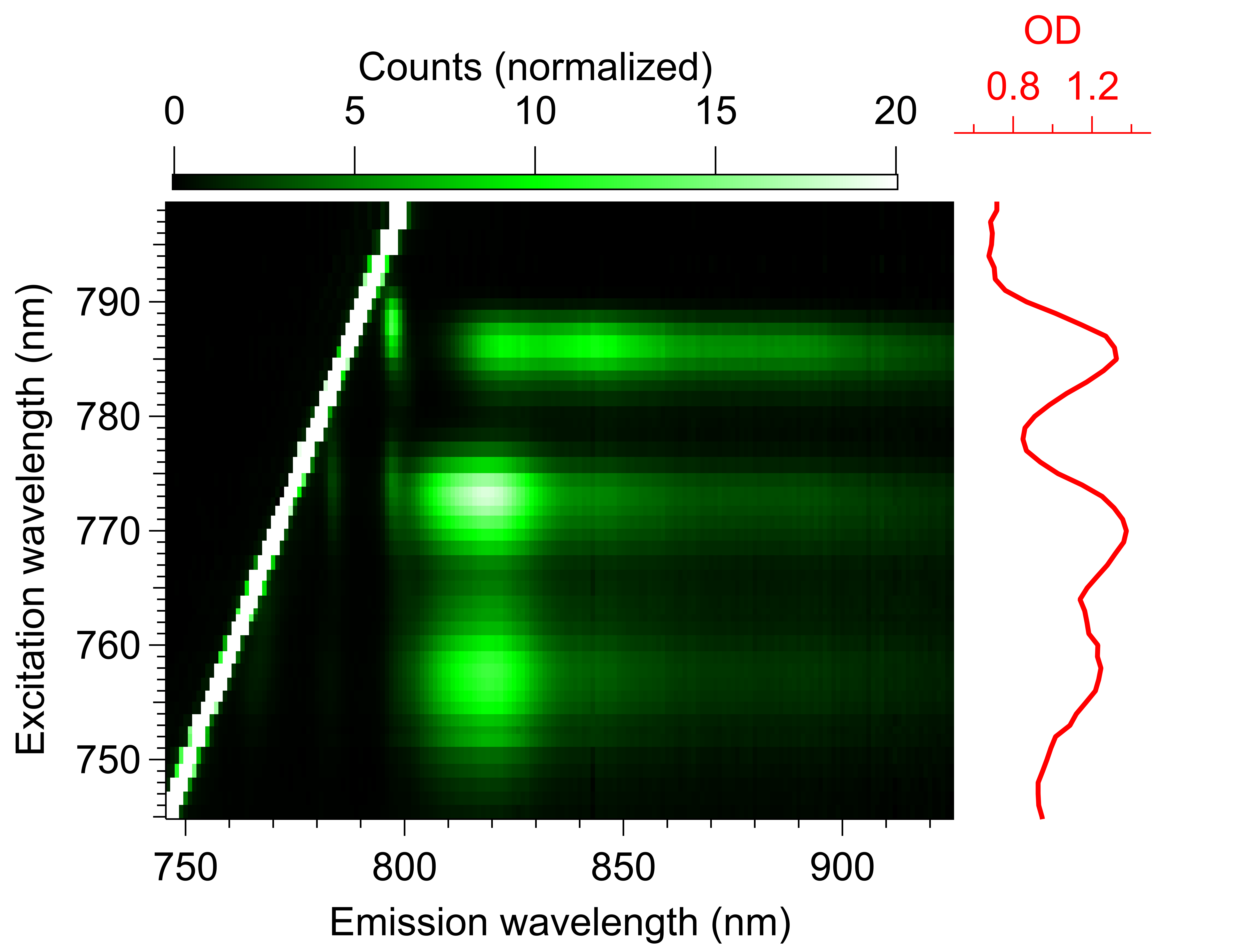}
    \caption{ 
{2D excitation-emission spectrum of Rb in Ne. As before, each emission spectrum has been normalized by the power of the excitation laser and the detector response, and the spectrometer is saturated by resonant scatter at the excitation wavelength. Included for reference at the right is the optical depth of the sample. 
}
\label{fig:Ne2Dspectrum}
    }
    \end{center}
\end{figure}

We measure Rb LIF spectra for laser excitation wavelengths from 740 to 925~nm, limited by the tuning range of our laser. This data is presented in Fig. \ref{fig:Ne2Dspectrum} as a 2D excitation-emission spectrum. No rubidium LIF signal was observed for excitation wavelengths longer than 800~nm, so that data was omitted from Fig. \ref{fig:Ne2Dspectrum}. For each of the three absorption peaks in the 740--800~nm range we see associated red-shifted fluorescence.

Comparing to prior work, we note that these LIF spectra are qualitatively  similar to what was observed for rubidium atoms trapped in the other noble gases. In the work of Gerhardt, Sin, and Momose  \cite{gerhardt2012excitation}, LIF was observed for rubidium-doped samples of argon, krypton, and xenon. They observed  fluorescence that was red shifted on the order of 100~nm, and emission linewidths were on the order of tens of nm (wider in argon than in krypton or xenon). They were unable to observe rubidium spectrum in neon films, but this was attributed to issues with the substrate temperature during growth. For the case of helium, attempts to observe LIF from Rb atoms trapped in different phases of solid helium reported either little or no LIF signal \cite{PhysRevLett.88.123002, PhysRevA.74.032509}. The observed fluorescence spectrum was qualitatively similar to argon, krypton, and xenon, but with additional red-shifted peaks that were attributed to the formation of rubidium-helium exciplexes.


For LIF measurements of single atoms, it is important to know not only the spectrum of the emitted light, but also the fraction of absorption events that result in the emission of a photon, which we refer to as the radiative quantum efficiency.
To measure this quantity, we image the LIF from the sample onto a silicon CMOS sensor. We excite the sample at 786~nm, and suppress the resonant scatter using a color filter and polarizer as described above. 
We calculate the rubidium radiative quantum efficiency using the measured light absorption, the collection efficiency of our optics (under the assumption that light is emitted isotropically), the transmission of our optics, and the specified detector quantum efficiency at the observed emission wavelengths. 
{
We find a radiative quantum efficiency of 6\%;
}
due to uncertainties in collection efficiencies, we expect this result is only accurate to within a factor of 3.

The measurements of the emitted light intensity reveal the \emph{ensemble average} of the probability that a photon absorption event results in the emission of a redshifted photon. Further insight into  \emph{single-atom} emission probabilities can be obtained through measurements of the excited-state lifetime of the atoms in the matrix.

\begin{figure}[ht]
    \begin{center}
    \includegraphics[width=\linewidth]{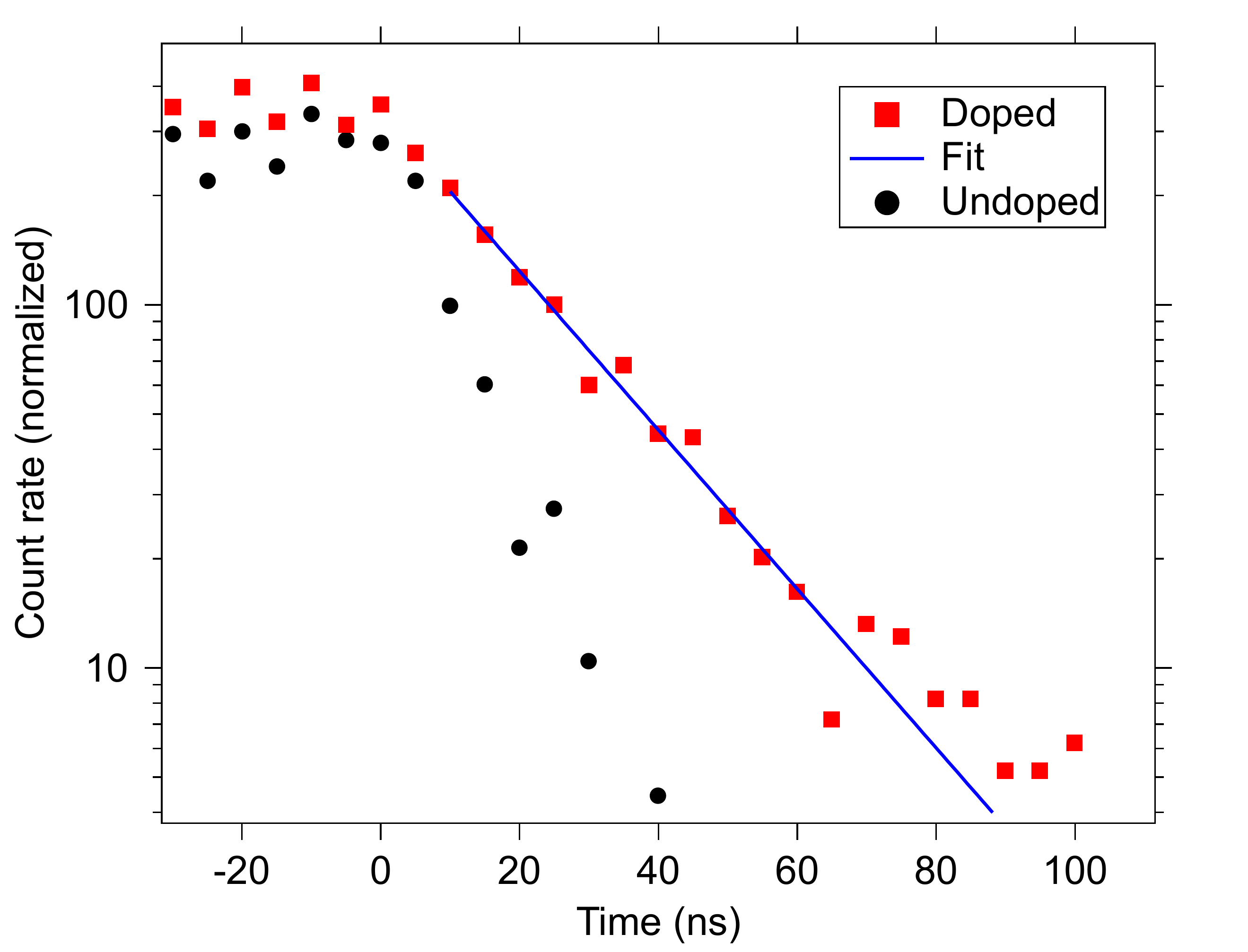}
    \caption{ 
LIF signal from Rb in Ne excited at 786~nm, as described in the text. For comparison, we also plot the signal from an undoped sample excited at the same wavelength.
The count rate from the undoped sample is much lower; both samples have been normalized to the same total number of counts.
The horizontal axis has been shifted so that $t=0$ coincides with the fall in the signal after the excitation light has been turned off, and the background count rate has been subtracted. The Rb LIF signal is fit to exponential decay.
\label{fig:NeLifetime}
    }
    \end{center}
\end{figure}

To measure the lifetime of the excited state, we rapidly modulate the intensity of our excitation laser with an acousto-optic modulator (AOM), and use a photon-counting photomultiplier tube to measure the photon count rate as a function of time after the light is turned off; a color filter is used to suppress resonantly-scattered light.
Sample data for excitation at 786~nm is shown in Fig. \ref{fig:NeLifetime}, for both doped and undoped samples. 
%
In the case of light emitted from an undoped sample we see elastic scattering from the neon matrix, and observe that the count rate falls on a timescale of $\sim7$~ns. We attribute this to the switching speed of our AOM. In the case of resonant excitation of a rubidium-doped sample, we see a significantly longer lifetime, indicative of the excited-state lifetime. 

We note that data for doped samples excited off-resonance shows similar behavior to the undoped sample, as does a resonantly-excited doped sample if the light is filtered to transmit resonant-scattered light and block red-shifted light.

For the red-shifted emission from the rubidium atoms, by fitting multiple data sets from two different samples 
we find an average decay time %
of 21.6~$\pm$~1.2~ns.  
Compensating for the finite shutoff time of our AOM and the response time of our electronics, we find --- for rubidium trapped in neon and excited at 786~nm --- an excited state lifetime of 19.7~$\pm$~1.3~ns. 
Whether this lifetime is uniform for all fluorescing rubidium atoms in the sample, or whether we are measuring an ensemble average of inhomogenous lifetimes is not known.


The natural lifetime of the $5p \ ^2P_{1/2}$ and  $5p \ ^2P_{3/2}$ excited states of an ``unperturbed'' rubidium atom are 28 and 26~ns, respectively \cite{NIST_ASD}. Because the lifetime in the matrix is only slightly shorter than the natural lifetime, we conclude that --- for the atoms which emit red-shifted light --- radiative decay is more likely than nonradiative quenching. 

However, the measurements of the intensity of the emitted light implies that a small fraction of the total number of excited-state atoms radiatively decay.
To reconcile these two observations, we speculate that there is inhomogenous behavior among our trapped atoms. A small fraction of the atoms that absorb at 786~nm are in  ``good'' trapping sites that radiatively decay with a high quantum efficiency, while the majority of rubidium atoms are ``dark'' due to their ``bad'' trapping sites, and decay predominantly through nonradiative channels.

This is similar to other solid-state systems proposed for use as quantum sensors
in which a fraction of the implanted ``defects'' have the desired properties \cite{barry2020sensitivity}.
For the creation of single-atom quantum sensors, this is an inconvenience but not a fatal flaw. It simply requires selecting an atom in a good trapping site; this selection process would be aided by the fact that the desired atoms are the only atoms which emit significant LIF. 



\section{LIF measurements: Parahydrogen}

\begin{figure}[ht]
    \begin{center}
    \includegraphics[width=\linewidth]{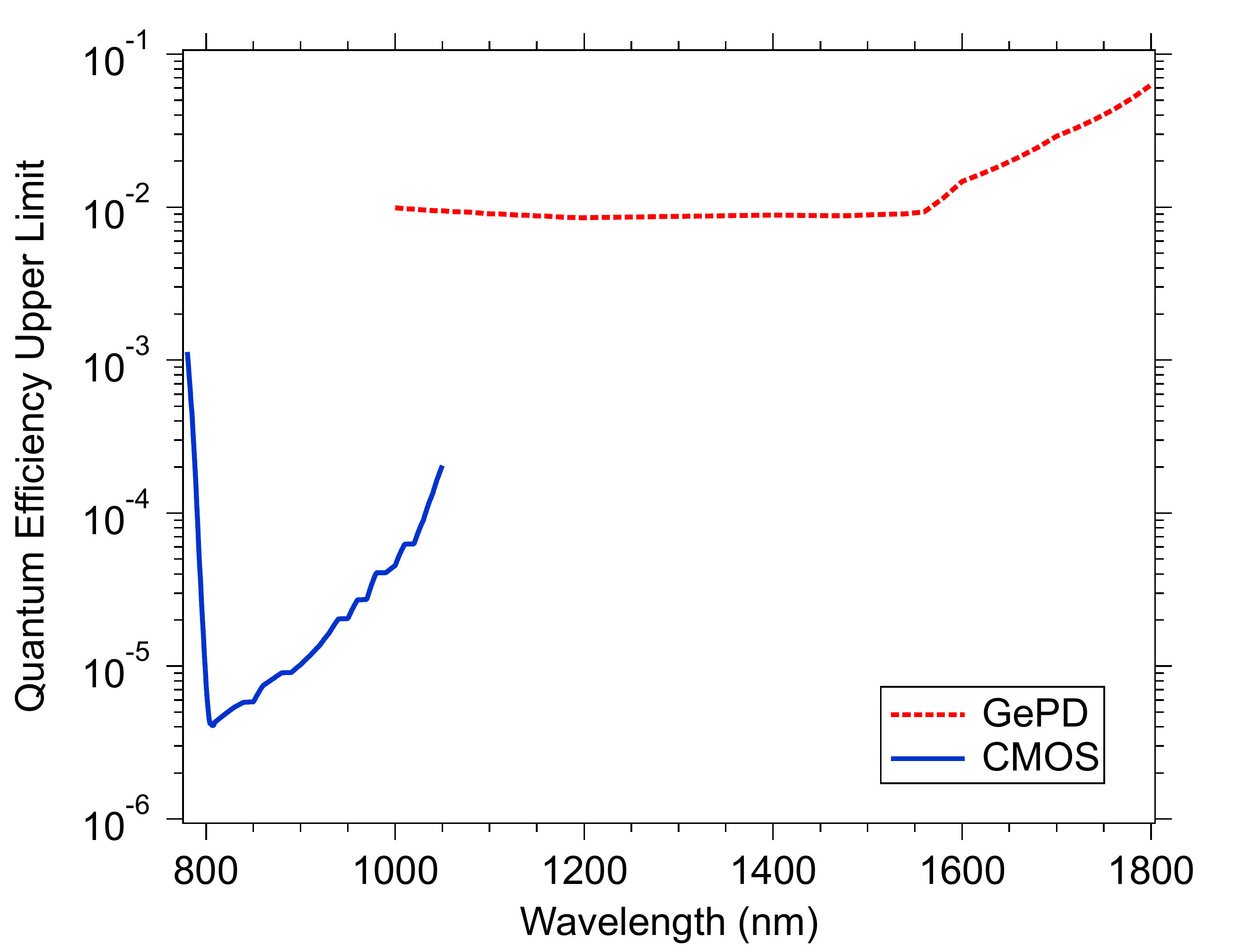}
    \caption{ 
The  upper limit on the radiative decay quantum efficiency for an ensemble of rubidium atoms in parahydrogen, as described in the text. The upper limit is obtained from measurements with a silicon CMOS detector and an amplified germanium photodiode, and is plotted as a function of the assumed emission wavelength.
\label{fig:H2QELimit}
    }
    \end{center}
\end{figure}

Rubidium-doped parahydrogen samples were grown as described in Refs. \cite{PhysRevA.100.063419, upadhyay2016longitudinal}, with similiar optical spectra as reported in those references, and as shown in the inset of Fig. \ref{fig:H2BleachingTime}.
We searched for laser-induced fluorescence by similar methods to our neon measurements, but did not observe LIF. From our measurements, we can put an upper limit on the radiative ``quantum efficiency'' of the ensemble of rubidium atoms: the ratio of emitted photons to absorbed photons from the excitation beam. 
This upper limit is shown in Fig. \ref{fig:H2QELimit}, and described below.

To obtain the limit shown in Fig. \ref{fig:H2QELimit}, we excite the sample with light from a narrow-linewidth laser at 755~nm, on resonance with the transition used for optical pumping and readout of spin \cite{PhysRevA.100.063419,upadhyay2016longitudinal}.
Emitted light from the sample is collected and focused into, in separate measurements, a fiber-coupled grating spectrometer, an imaging silicon CMOS detector, and (to look further into the infrared) a germanium photodetector. The large amount of elastically scattered light from the matrix makes alignment straightforward.

To search for red-shifted LIF, we use a long-pass interference filter to block the 
resonantly-scattered light from the matrix.
In the case of the silicon detector, light from the sample was observed. To determine whether this was red-shifted light or simply ``leakage'' of resonantly-scattered light through the filter, we removed the filter from the detector and placed it in the optical path prior to the sample. This arrangement gave a comparable signal, indicating the signal was predominantly from leakage of resonant light. 
We use the measured light levels to provide an upper limit on the radiative quantum efficiency, performing the calculation in a manner similar to what was done for the signal of rubidium in neon. In the case of the germanium photodetector --- which has a much higher noise background --- no LIF signal was observed above the detector noise floor, which was similarly used to set an upper limit.
The upper limit shown in Fig. \ref{fig:H2QELimit} is plotted under the assumption that the rubidium fluorescence is emitted at a single wavelength.
{
Because the detector response and the transmission of the interference filter both depend on the frequency of the detected light, the upper limit depends on the assumed emission wavelength.} 

Exciting the sample at 726~nm (the shoulder of the neighboring absorption peak) also produced no observable LIF, with upper limits comparable to what is shown in Fig. \ref{fig:H2QELimit}.

From prior work with matrix-isolated alkali atoms, we expect any fluorescence produced to be red-shifted by tens to hundreds of nanometers. However, we also attempted to observe resonantly-scattered light (as well as light scattered at wavelengths close to resonance) from rubidium atoms by measuring the emitted light without filters.
%
Because of the large background scatter from the matrix, these measurements are considerably less sensitive than the measurements of red-shifted light.
When we excite a doped sample on resonance with the rubidium transition, we see \emph{less} emitted light than if we excite an undoped sample at the same wavelength, or excite a doped sample with off-resonance light.
This is as expected for a poor rubidium radiative quantum efficiency: for an undoped sample (or a doped sample excited off-resonance), we detect the unattenuated scatter from the matrix. If we excite a doped-sample on-resonance, the rubidium atoms attenuate both the excitation light and the resonant scattered light from the matrix. We construct a simple model for this process, and from the observations of the relative intensities scattered by samples on- and off-resonance, we put an upper limit on the rubidium quantum efficiency of 0.1.


We note we can also place a lower limit on the quantum efficiency of the trapped atoms from their spectrum. The measured linewidth places a upper limit on the \emph{total} decay rate of the excited state from the time-energy uncertainty relation. Assuming the \emph{radiative} decay rate is not changed significantly from its vacuum value --- reasonable for a weakly-interacting host matrix such as parahydrogen --- we can find a lower limit on the quantum efficiency from the ratio of the radiative decay rate and (the upper limit of) the total decay rate. From this, we find the quantum efficiency is $\geq 10^{-7}$.






\section{Bleaching}

Equally important for the development of practical quantum sensors is the phenomenon of ``bleaching''. While gas-phase atoms can be optically cycled an unlimited number of times without any change to their electronic structure or absorption spectrum, the optical excitation of atoms trapped in the solid phase can cause changes in their optical properties. We refer to this effect as bleaching.

Bleaching was previously observed for potassium atoms in parahydrogen: after absorbing roughly $10^4$ photons, atoms would no longer absorb light at that wavelength \cite{PhysRevA.100.063419}. The loss of absorption at the original wavelength was accompanied by an increase in absorption at other wavelengths. These changes in the spectrum were attributed to population transfer between different trapping sites in the matrix caused by electronic excitation \cite{PhysRevA.100.063419}. For potassium in parahydrogen, no method of reversing these changes was found.

Similar observations were made for alkali atoms in argon \cite{gerhardt2012excitation}. In argon it was observed that bleaching would occur after many fewer absorption events ($\sim 10$), but the process could be reversed by the application of light at the wavelengths that saw increased absorption after bleaching. The interpretation was that the  excitation at this second wavelength which would return the atoms to their original trapping site \cite{kanagin2013optical}.

In this work, we find rubidium atoms in parahydrogen are more resistant to bleaching than either of these prior results, and rubidium atoms trapped in neon are more resistant still.

\begin{figure}[ht]
    \begin{center}
    \includegraphics[width=\linewidth]{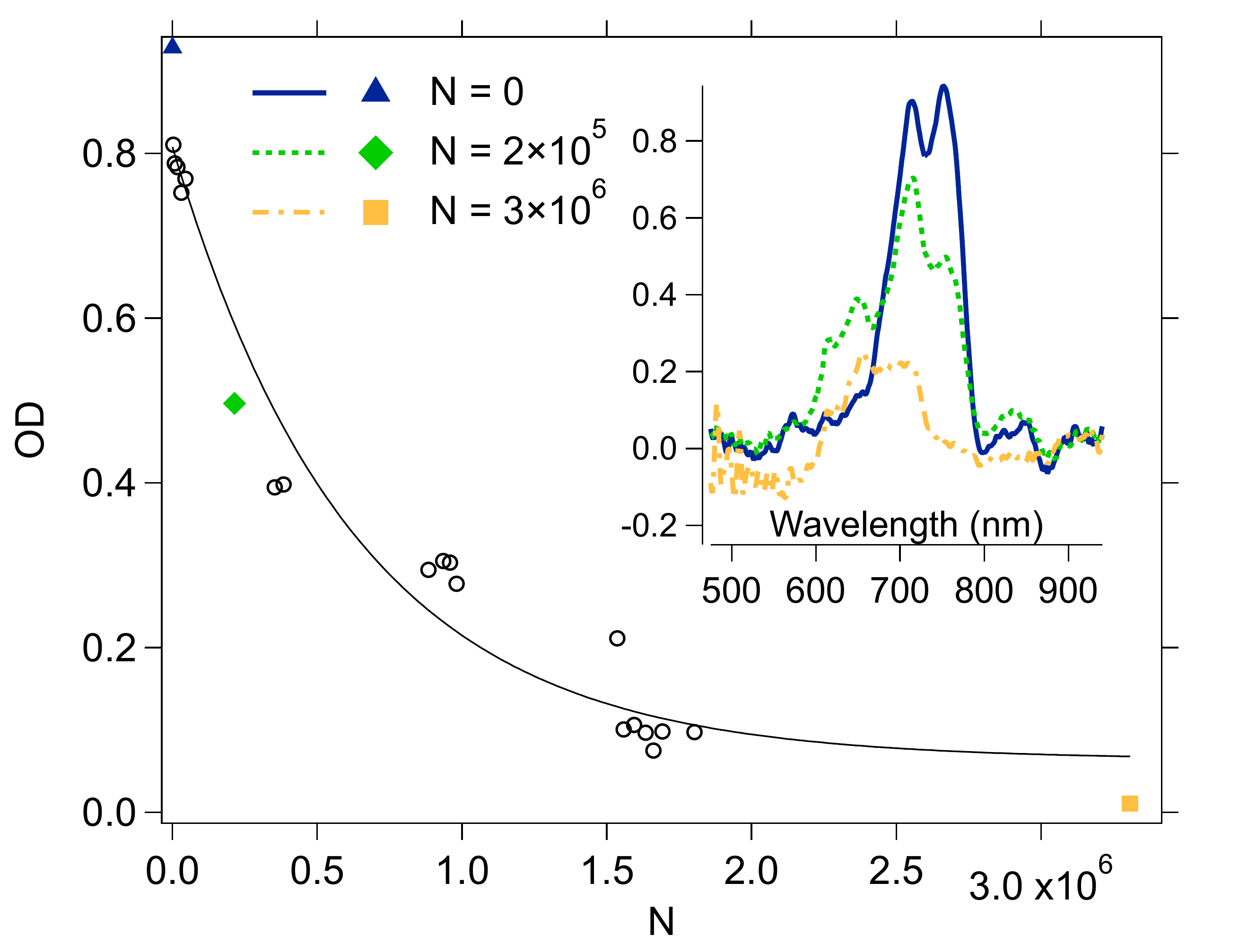}
    \caption{ 
The main graph shows the background-subtracted OD at 754~nm as a function of the sample exposure 
expressed as $N$, the number of photons absorbed per unbleached atom (as described in the text).
The data is fit to exponential decay.
The inset shows three spectra taken at different stages of the bleaching; the corresponding points on the main graph are shown as filled symbols.}
\label{fig:H2BleachingTime}
    \end{center}
\end{figure}

Data for the bleaching of rubidium trapped in parahydrogen is shown in Fig. \ref{fig:H2BleachingTime}. In this data, the entire sample was exposed to 754~nm laser light, with a beam diameter $\sim 1$~cm, and laser powers from $\sim3$ to $\sim 20$~mW.
During bleaching, we measure the optical depth using white-light absorption spectroscopy. We subtract the background absorption signal from the parahydrogen matrix under the assumption that it is equal to a linear interpolation of the measured off-resonance optical depths at 516 and 887~nm.

As seen in Fig. \ref{fig:H2BleachingTime}, the spectrum undergoes dramatic changes due to exposure to light, including both the reduction of the optical depth of the feature being excited (and, eventually, its neighbor) and the increase of the optical depth at other wavelengths.

We plot the background-subtracted OD at different points in the bleaching test and characterize the bleaching time as a number of photons absorbed per unbleached atom ($N$), as shown in Fig. \ref{fig:H2BleachingTime}. We calculate the rate at which photons are absorbed by rubidium atoms from the beam power and the background-subtracted optical depth. We calculate the number of unbleached atoms from the measured optical depth and the pre-bleaching linewidth. From the combination of these two, we find $N$. Fitting the data to a single exponential, yields a bleaching ``lifetime'' of 6$\times 10^5$~photon absorption events, with a factor of 2 uncertainty. 
 
Rubidium atoms in neon are significantly more resistant to bleaching. With a large-diameter beam addressing the entire sample, we observed no changes in the white-light absorption spectrum. To search for bleaching, we use a more tightly focussed beam, and observe the transmission of that beam as a function of time: we exposed the sample to 785~nm laser light with powers of 0.2 to 5~mW 
and a beam diameter of $\sim$~310~$\mu$m. 
{
After the absorption of 10$^9$ photons per rubidium atom, the optical depth of the sample at the excitation wavelength decreased by $(17 \pm 3)\%$ and the LIF signal decreased by $(32\pm 4)\%$.
These numbers were consistent across multiple samples (to within the stated error bars) and suggest that the atoms with high radiative quantum efficiency are slightly more susceptible to bleaching.
}

\section{Conclusions}

For rubidium atoms trapped in solid parahydrogen, we were unable to observe laser-induced fluorescence, and placed an upper limit on the quantum efficiency of emission. We note that the measured upper limit is for the ensemble of atoms; it is possible that a small fraction of the trapped atoms radiate efficiently. However, with no evidence for LIF,  rubidium atoms in parahydrogen do not appear a promising choice for single-atom quantum sensors.

We note that prior work with boron atoms trapped in solid parahydrogen observed an LIF signal (with a linewidth narrower than seen with alkali-metal atoms in noble gas matrices) \cite{tam2000electronic}. LIF has also been observed from molecules trapped in parahydrogen \cite{huang2016laser, tsuge2020spectroscopy}. So rubidium's lack of LIF is not universal. Whether or not it is universal for all alkali-metal atoms is unknown. Differing behavior between alkali atoms has been observed before in solid and liquid helium, in which some alkali atoms exhibited strong LIF signals, while others did not \cite{moroshkin2006spectroscopy}. However, we are not optimistic. Similar to our rubidium measurements, Fajardo and Tam reported laser-induced fluorescence was observed for sodium atoms trapped in solid neon, but not for sodium atoms trapped in parahydrogen \cite{fajardo1998solid}. 

Rubidium in solid neon exhibits much better optical properties. A red-shifted LIF signal was observed, and it was determined that a small fraction of the trapped rubidium atoms are efficient optical emitters. Rubidium in solid neon can be optically cycled many times ($\gtrsim 10^9$ times) without bleaching.  These properties are promising for using single rubidium atoms trapped in solid neon as single-atom quantum sensors. However, other key requirements for an electron spin-based quantum sensor --- such as the ability to optically pump and measure the spin state of the trapped atoms and the ability to achieve long electron spin coherence times --- are unknown, and merit investigation. 
{
From prior work measuring alkali atoms trapped in the hcp phase of solid helium \cite{moroshkin2006spectroscopy}, argon \cite{kanagin2013optical}, and parahydrogen \cite{PhysRevA.100.063419} we are cautiously optimistic; 
we plan to measure these properties in future work. If those properties are as favorable as expected, we plan to transition to single-atom work using LIF, using similar methods as reference \cite{chambers2019imaging}.
}

\section{Acknowledgements}

This material is based upon work supported by the National Science Foundation under Grant No. PHY-1912425.
We gratefully acknowledge helpful conversations with David Patterson and  David Todd Anderson.

\bibliography{Rb_H2_LIF_2020.bib}

\end{document}